\documentclass[11pt]{article}
\usepackage{comment}
\usepackage[margin=2.5cm]{geometry}
\usepackage{amsmath,amssymb,extarrows,mathtools,graphicx,subfigure,setspace,bbold,physics}
\usepackage{cite}
\usepackage{slashed}
\usepackage[dvipsnames]{xcolor}
\usepackage{hyperref,caption,diagbox}
\hypersetup{colorlinks=true, linkcolor=blue, citecolor=magenta, linktoc=page}
\usepackage{tikz}
\usepackage{color}
\usepackage{arydshln,leftidx,mathtools}
\usetikzlibrary{decorations.pathreplacing}
\numberwithin{equation}{section}
\newcommand{\be}{\begin{equation}}
	\newcommand{\bea}{\begin{eqnarray}}
		\newcommand{\eea}{\end{eqnarray}}
	\newcommand{\ba}{\begin{align}}
		\newcommand{\ea}{\end{align}}
	\newcommand{\ee}{\end{equation}}

\usepackage{chngcntr}
\begin{document}
\begin{titlepage}
		\thispagestyle{empty}

		\begin{flushright}
			IPM/P-2023/53\\
		\end{flushright}
			\vspace{.4cm}
	
	\vspace{.4cm}
	\begin{center}
		\noindent{\Large \textbf{ Krylov Complexity in Lifshitz-type Scalar Field Theories}}\\
		
		\vspace*{15mm}
		\vspace*{1mm}
		{M. J. Vasli$^a$,  K. Babaei Velni$^{a,c}$, 
  M. R. Mohammadi Mozaffar$^a$, 
  A. Mollabashi$^b$ 
  and  M. Alishahiha$^c$
}
		
		\vspace*{1cm}
		
		{\it $^a$ Department of Physics, University of Guilan, P.O. Box 41335-1914, Rasht, Iran\\
  $^b$ Center for Gravitational Physics, Yukawa Institute for Theoretical Physics, Kyoto University,
Kitashirakawa Oiwakecho, Sakyo-ku, Kyoto 606-8502, Japan\\
$^c$ School of Physics, Institute for Research in Fundamental Sciences (IPM), P.O.Box 19395-5531, Tehran, Iran
		}
		
		\vspace*{0.5cm}
		{E-mails: {\tt vasli@phd.guilan.ac.ir,\{babaeivelni, mmohammadi\}@guilan.ac.ir, ali.mollabashi@yukawa.kyoto-u.ac.jp,
 alishah@ipm.ir}}%
		
		\vspace*{1cm}
		%%\maketitle
	\end{center}

\begin{abstract}
We investigate various aspects of the Lanczos coefficients in a family of free Lifshitz scalar theories, characterized by their integer dynamical exponent, at finite temperature. In this non-relativistic setup, we examine the effects of mass, finite ultraviolet cutoff, and finite lattice spacing on the behavior of the Lanczos coefficients. We also investigate the effect of the dynamical exponent on the asymptotic behavior of the Lanczos coefficients, which show a universal scaling behavior. We carefully examine how these results can affect different measures in Krylov space, including Krylov complexity and entropy. Remarkably, we find that our results are similar to those previously observed in the literature for relativistic theories.
\end{abstract}
		
\end{titlepage}

	\newpage
	
	\tableofcontents
	\noindent
	\hrulefill
	
	\onehalfspacing
%%%%%%%%%%%%%%%%%%%%%%%%%%

\section{Introduction}

Quantum chaos is an interesting subject though it is difficult
to understand. This is due to the fact that the time evolution 
of quantum mechanics is local and unitary and thus, in general, it is hard to study the emergence of ergodic behavior in quantum systems. Therefore it is of great interest to understand thermal behavior at the quantum level in which the eigenstate thermalization hypothesis plays an important role \cite{{Deutsch:1991},{Srednicki:1994mfb},{DAlessio:2015qtq},{Srednicki:1999}}.

At the classical level, the chaotic behavior may be associated with the sensitivity of trajectories in the phase space to the initial conditions. Indeed, in chaotic systems two initially nearby trajectories separate exponentially fast characterized by the Lyapunov exponent. Having dealt with trajectories, it is then clear why the generalization of chaos to the quantum level should not be straightforward.

Nonetheless, to probe the nature of quantum chaos certain quantities have been introduced in the literature. These include, for example, out-of-time-order correlators (OTOCs)  \cite{{Shenker:2013yza},{Shenker:2013pqa}}. Semiclassically due to the butterfly effect, OTOCs exhibit an exponential growth characterized by the Lyapunov exponent which is conjectured to be bounded \cite{Maldacena:2015waa}. The bound saturates for certain strongly interacting models which have holographic descriptions such as the Sachdev-Ye-Kitaev (SYK) model \cite{{Kitaev},{Sachdev:1993},{Maldacena:2016hyu}}. We note, however, that the exponential growth of OTOCs is not a generic feature of chaotic systems\cite{{Fine},{Xu:2018xfz}}.

It is an interesting problem to explore a possible uniform approach that could describe the chaotic nature of chaotic systems. Recently, it has been proposed that the operator growth in many body systems may have enough information to make a distinction between chaotic and non-chaotic systems \cite{Parker:2019}. In a general many-body quantum system, the evolution of an operator is given by the Heisenberg equation, ${\cal O}(t)=e^{-iHt}{\cal O}e^{iHt}$, by which a simple operator may become rather a complex operator as time evolves. Here $H$ is the Hamiltonian of the system. More precisely, at any time the operator may be expanded in terms of nested operators $[H,[H,\cdots,[H,{\cal O}]]]$ as follows
\be\label{timeevolution}
{\cal O}(t)=e^{-iHt}{\cal O}e^{iHt}=\sum_{n=0}^\infty \frac{(it)^n}{n!}{\cal L}_n\,,
\ee
where ${\cal L}_n=\{{\cal O}, [H,{\cal O}], [H,[H,{\cal O}]],\cdots\}$. These nested operators, given a proper inner product in the space of operators,  are not orthogonal and normalized, although it is possible to construct an orthogonal and ordered basis known as the Krylov basis. The procedure by which the basis is constructed is known as the Gram-Schmidt process. The construction of the  Krylov basis via a recursion method amounts to defining Lanszoc coefficients $b_n$ that contain information on the dynamics of the operator in the Krylov space. It is then natural to define a measure that probes the growth of the operator over the Krylov basis which could be thought of as an indicator of how complex the operator might become as time evolves. The corresponding measure is known as the Krylov complexity or K-complexity, which has been the subject of a wide regain interest  from  many body systems to holography
\cite{{Barbon:2019wsy},{Avdoshkin:2019trj},{Dymarsky:2019elm},{Rabinovici:2020ryf},{Cao:2020zls},{Dymarsky:2021bjq},{Trigueros:2021rwj},{Rabinovici:2021qqt},{Fan:2022xaa},{Hornedal:2022pkc},{Balasubramanian:2022tpr},{Heveling:2022hth},{Bhattacharjee:2022vlt},{Adhikari:2022oxr},{Adhikari:2022whf},{Caputa:2022eye},{Muck:2022xfc},{Bhattacharya:2022gbz},{Bhattacharjee:2022qjw},{Afrasiar:2022efk},{Alishahiha:2022nhe},{He:2022ryk},{Bhattacharjee:2022ave},{Bhattacharjee:2022lzy},{Alishahiha:2022anw},{Bhattacharya:2023zqt},{Erdmenger:2023shk},{Pal:2023yik},{Nizami:2023dkf},{Rabinovici:2023yex},{Nandy:2023brt},{Patramanis:2023cwz},{Bhattacharyya:2023dhp},{Camargo:2023eev},{Caputa:2023vyr}}.

%(K-complexity) that is a measure of %the operator growth
%over the Krylov space  

The authors of \cite{Parker:2019} proposed a universal operator growth hypothesis that relates the asymptotic behavior of the Lanczos coefficients to the nature of the dynamics of the system under consideration. More precisely, for a chaotic many-body quantum system (for dimensions greater than one) without symmetry the Lanczos coefficients, asymptotically, should grow linearly 
\bea\label{bnfit}
b_n=\alpha n+\gamma + O(1),\,\,\,\,\,\,\,\;\;\;\;\;\;\;\;\;\;\;{\rm for}
\,\,\;\;\;n\gg 1,
\eea
where $\alpha>0$ is a real constant referred as the growth rate and $\gamma$ is also a constant. In this case, the K-complexity exhibits an exponential growth with an exponent $\lambda=2\alpha$. This hypothesis is motivated by the behavior of the power spectrum (that is the Fourier transformation of auto-correlation function) at high frequency limit \cite{Elsayed:2014}. Indeed the linear growth of the Lanczos coefficients is equivalent to the exponential decay of the power spectrum which in turn is equivalent to a pole in the auto-correlation function. 
  
Although, for chaotic systems considered in the literature, the Lanczos coefficients exhibit asymptotic linear growth, it seems that the above proposal is not universal in the sense that the linear growth may not be directly related to the chaotic nature of the system. Indeed, the asymptotic linear growth may also occur even in non-chaotic models \cite{{Dymarsky:2021bjq},{Bhattacharjee:2022vlt}}. Actually, for continuous systems such as quantum field theory, the situation is even worst in the sense that for a local operator the Lancsoz coefficients always exhibit linear growth unless we add extra ingredients to the system, such as adding a hard cutoff or putting the theory on a compact space \cite{{Avdoshkin:2022xuw},{Camargo:2022rnt}}. Intuitively, this is because, for any field theory, the singularity of the two-point function when the operators approach each other yields an exponential decay in the power spectrum, which automatically results in  an asymptotic linear growth for Lanczos coefficients, preventing it from being a good probe for chaos. 

As we mentioned the asymptotic linear growth of the Lanczos coefficients implies an exponential growth of the K-complexity in the asymptotic limit $t\rightarrow \infty$, i.e, 
\begin{eqnarray}\label{KCexp}
K_{\mathcal{O}}(t)\propto e^{\lambda_K t},
\end{eqnarray}
where the exponent $\lambda_K$ controls the rate of change of $K_{\mathcal{O}}(t)$. It was shown in \cite{Parker:2019} that in local quantum many-body systems at infinite temperature  with finite-dimensional Hilbert spaces, $\lambda_K$ bounds the Lyapunov exponent, i.e., $\lambda_L\leq\lambda_K$, which conjecturally applies even at finite temperature, which would put even tighter bound on chaos. Moreover, from the general behavior of the exponent $\lambda_K$ for systems at finite temperature, it is plausible to conjecture the following inequality  \cite{Avdoshkin:2019trj}
\begin{eqnarray}\label{lambdaKbound}
\lambda_L\leq \lambda_K\leq 2\pi T.
\end{eqnarray}
The aim of this article is to further explore the behavior of Lanczos coefficients, K-complexity, and the above conjectural bound for certain systems with Lifshitz scaling symmetry acting as
\be
t\rightarrow \lambda^z t,\;\;\;\;\;\;\;\;x_i\rightarrow \lambda x_i,
\ee
where $t$ is time and $x_i$'s are spatial directions of the space-time. Moreover, $z$ denotes the dynamical critical exponent that determines the anisotropy between time and space such that for $z=1$ the relativistic scaling is recovered. A quantum field theory that respects the above symmetry is a Lifshitz field theory\footnote{Further discussions on
different aspects of Lifshitz symmetries can be found in e.g. \cite{Hartong:2015wxa, Hartnoll:2016apf, Figueroa-OFarrill:2022kcd}.} (see \cite{Alexandre:2011kr} for a review). In particular, we consider a $d$-dimensional scalar theory which is a generalization of relativistic Klein-Gordon theory and respects Lifshitz scaling symmetry in the massless limit with the following action \cite{Alexandre:2011kr}
\begin{eqnarray}\label{Lifaction}
S=\frac{1}{2}\int dt\, d^{d-1}x \left(\dot{\phi}^2-\sum_{i=1}^d \left(\partial_i^z \phi\right)^2+m^{2z} \phi^2\right),
\end{eqnarray}
where the dot indicates derivative with respect to $t$. The corresponding dispersion relation takes the form
\begin{eqnarray}\label{Lifdispersion}
\epsilon_k^2=k^{2z}+m^{2z},
\end{eqnarray}
where $k^{2z}=\sum_{i=1}^{d} k_i^{2z}$. By replacing the space continuum with a discrete mesh of lattice points the above expression can be transformed into a discrete counterpart as follows \cite{MohammadiMozaffar:2017nri}
\begin{eqnarray}\label{Lifdispersionlattice}
\epsilon_k^2=\sum_{i=1}^{d-1}\left(2\sin\frac{\pi k_i}{N}\right)^{2z}+m^{2z},
\end{eqnarray}
where we assume a hypercubic lattice with length $N$ in every spatial direction. Recently, there have been many attempts to investigate various properties of information measures, including entanglement entropy, in such theories. These investigations have led to a remarkably rich and varied range of new insights, e.g., \cite{MohammadiMozaffar:2017nri, He:2017wla, Gentle:2017ywk, MohammadiMozaffar:2017chk, MohammadiMozaffar:2018vmk, MohammadiMozaffar:2019gpn, Hartmann:2021vrt, Mozaffar:2021nex, Mintchev:2022xqh, Mintchev:2022yuo}. Related investigations attempting to better understand quantum chaos, computational complexity, and entanglement measures in the context of Lifshitz holography have also been reported in \cite{Alishahiha:2014cwa, Roberts:2016wdl, Alishahiha:2018tep}. 

The remainder of our paper is organized as follows: In
Sec. \ref{review}, we give the general framework in which we are working, establishing our notation and the general form of the Lanczos coefficients, K-complexity, and other related quantities in the Krylov space. In Sec. \ref{Lifshitz scalar theory}, we consider the continuum case and study the properties of Lanczos coefficients and K-complexity numerically. We present a combination of numerical and analytic results on the scaling of these quantities. To get a better understanding of the  results, we will also compare the behavior of complexity to other measures including K-entropy. In Sec. \ref{UVCUT}, we extend our studies in the presence of a UV cutoff, either by introducing a finite UV cutoff in continuous momentum space or considering a discretized version of our model with finite lattice spacing. We review our main results and discuss their physical implications in Sec. \ref{Conclusions}, where we also indicate
some future directions.

\section{A brief review of Lanczos algorithm}\label{review}

In this section, we employ the Lanczos algorithm to find the Lanczos coefficients, by which we may compute several interesting quantities, such as K-complexity. As we already mentioned, in order to study operator growth in the Krylov space, one needs to define a proper inner product. Since we are interested in a system at a finite temperature, the appropriate inner product may be defined by the Wightman inner product
\begin{eqnarray}\label{innernotation}
(\mathcal{O}|\mathcal{O}')\equiv \langle e^{\frac{\beta H}{2}} \mathcal{O}^\dagger e^{-\frac{\beta H}{2}}\mathcal{O}'\rangle_\beta.
\end{eqnarray}
Using this inner product, one could construct the Krylov space starting from an initial operator ${\cal O}(0)$. Denoting the Krylov basis by $\{|{\cal O}_n)\}$, the evolved operator at a given time may be expanded in this basis as follows
\begin{eqnarray}\label{AU}
|\mathcal{O}(t))=\sum_{n=0}^{\infty} i^n\phi_n(t)|\mathcal{O}_n),
\end{eqnarray}
where due to the normalization condition we have $\sum_{n=0}^{\infty} |\phi_n(t)|^2=1$. The probability amplitudes $\phi_n(t)$ may be computed recursively from the following Schr\"odinger equation 
\begin{eqnarray}\label{disSchrodinger}
\frac{d\phi_n}{dt}=b_n\phi_{n-1}-b_{n+1}\phi_{n+1},
\end{eqnarray}
with the boundary conditions $\phi_n(0)=\delta_{n0},\,\phi_{-1}(t)\equiv 0$. 

Having found the probability amplitudes $\phi_n(t)$, one may define several physical quantities that could probe the nature of the operator growth, which in turn could give us information about the nature of the dynamics of the system under study. The most famous quantity in this context is the K-complexity, defined by
\begin{eqnarray}\label{KC}
K_{\mathcal{O}}(t)=\sum_{n=0}^{\infty}n |\phi_n(t)|^2.
\end{eqnarray}
Moreover, motivated by \cite{Caputa:2021ori}, in order to gain a better insight into the properties of $K_{\mathcal{O}}(t)$, one may also define the $k$-th order K-variance 
\begin{eqnarray}\label{Kvariance}
\delta_{\mathcal{O}}(t)=\frac{\left(\sum_{n} n^k |\phi_n(t)|^2-K_{\mathcal{O}}(t)^k\right)^{\frac{1}{k}}}{K_{\mathcal{O}}(t)},
\end{eqnarray}
which measures the fluctuations around the average. Of course, in the present paper, we will mainly consider the case of $k=2$.

One can also extract further properties of the distribution of probability amplitudes $\phi_{n}(t)$ by studying entropic measures such as operator entropy, or K-entropy, which is defined through the von Neumann entropy of the probabilities as follows \cite{Barbon:2019wsy}\footnote{More generically, one can compute Renyi K-entropies, which is defined through the moments of the probability distribution, see e.g., \cite{Caputa:2021sib}.}
\begin{eqnarray}\label{Kentropy}
S_{\mathcal{O}}(t)=-\sum_{n}|\phi_n(t)|^2\;\log |\phi_n(t)|^2.
\end{eqnarray}
Clearly, if the amplitudes are very peaked at a particular value of $n$, the K-entropy is small, while for uniform distribution, it becomes large.

From the definition of the above quantities, we see that they can be computed if we know the explicit form of the probability amplitudes. Of course, to obtain the probability amplitudes, one needs to know the explicit form of the Lanczos coefficients. Therefore, one may conclude that all information about the operator growth is, indeed, encoded in the Lanczos coefficients. Thus the aim is to see how these coefficients can be computed for a given system. In order to find the Lanczos coefficients $b_n$ it is convenient to define the moments $\{\mu_{2n}\}$
\begin{eqnarray}\label{mu2n}
\mu_{2n}=
%(-i)^{2n}\frac{d^{2n}\Pi^W(t)}{dt^{2n}}\bigg|_{t=0}=
\frac{1}{2\pi} \int_{-\infty}^{\infty}d\omega \;\omega^{2n} f(\omega),
\end{eqnarray}
where $f$ is the power spectrum which is defined as the Fourier transformation of auto-correlation function, i.e., $\phi_0(t)=(\mathcal{O}(t)|\mathcal{O}(0))$, as follows
\begin{eqnarray}\label{Ftrans}
f(\omega)=\int_{-\infty}^{\infty}d\omega \;e^{i\omega t}\phi_0(t)\,.
\end{eqnarray}
Indeed, the moments $\mu_{2n}$'s are the Maclaurin expansion coefficients of the auto-correlation function. As shown in \cite{Viswanath:1994}, having known the moments, the Lanczos coefficients can be computed using the following recursion relation
\begin{eqnarray}\label{recursion}
&&b_{n}=\sqrt{M_{2n}^{(n)}},\hspace*{1cm} b_{-1}=b_{0}\equiv 1,\hspace*{1cm}M_{2\ell}^{(-1)}=0,\hspace*{1cm}M_{2\ell}^{(0)}=\mu_{2\ell},\nonumber\\
&&M_{2\ell}^{(j)}=\frac{M_{2\ell}^{(j-1)}}{b_{j-1}^{2}}-\frac{M_{2\ell-2}^{(j-2)}}{b_{j-2}^{2}},\hspace*{2cm}\ell=j,\cdots, n.
\end{eqnarray}

Let us emphasize that, in general, it is not possible to find a closed-form expression for $\phi_n(t)$, and hence finding the full-time profile of the quantities defined above  requires some numerical treatment. This amounts to modifying, for example, the Eq. \eqref{KC} and the normalization condition as follows
\begin{eqnarray}\label{KCapprox}
K_{\mathcal{O}}(t)\approx\sum_{n=0}^{n_{\rm max}}n |\phi_n(t)|^2, \hspace*{1cm}{\rm with}\hspace*{1cm}\sum_{n=0}^{n_{\rm max}} |\phi_n(t)|^2\approx 1.
\end{eqnarray}
This means that, actually, we are approximately computing K-complexity using the above equation for some finite $n_{\rm max}$.

Now we are equipped with all we need to study the behavior of Lanczos coefficients and thereby other quantities defined in this section for quantum field theories with the Lifshitz  symmetry.
%QFTs. Before examining the full dependence of these quantities on %different parameters, we would like to study the $z$-dependence of %them in certain cases including small and large mass regimes. As we %will see in some specific regimes the moments can be obtained %analytically. 

\section{Lanczos coefficients and Krylov complexity in Lifshitz scalar theory}\label{Lifshitz scalar theory}

In this section, following \cite{{Avdoshkin:2022xuw},{Camargo:2022rnt}}, we would like to find the Lanczos coefficients for the model introduced in equation \eqref{Lifaction} at finite temperature, which can be used to compute the quantities we introduced in the previous section.

To start, let us consider the thermal Wightman two-point function, also known as the auto-correlation function, which is 
\begin{eqnarray}\label{Pi}
\Pi^{W}(t)=\phi_0(t)=(\mathcal{O}(t)|\mathcal{O}(0))\equiv \langle \mathcal{O}^\dagger\left(t-\frac{i\beta}{2}\right)\mathcal{O}(0)\rangle_\beta,\hspace*{1cm}{\rm with}\hspace*{1cm}\beta=\frac{1}{T}\,.
\end{eqnarray}
Then the Wightman power spectrum, $f^W(\omega)$,  can be expressed in terms of the spectral function $\rho(\omega, k)$ as follows
\begin{eqnarray}\label{fW}
f^W(\omega)=\int_{-\infty}^{\infty}d\omega \;e^{i\omega t}\Pi^{W}(t)=\frac{1}{\sinh \frac{\beta \omega}{2}}\int \frac{d^{d-1}k}{(2\pi)^{d-1}}\rho(\omega, \vec{k}).
\end{eqnarray}
The spectral function $\rho(\omega, \vec{k})$ for the free massive scalar theory is given by \cite{Laine:2016hma}
\begin{eqnarray}\label{rho}
\rho(\omega,\vec{k})=\frac{\mathcal{N}}{\epsilon_k}\left(\delta(\omega-\epsilon_k)-\delta(\omega+\epsilon_k)\right)\,.
\end{eqnarray}
Here $\mathcal{N}$ is a normalization factor and $\epsilon_k$ denotes the energy eigenvalues. 
In our case where we are dealing with the free Lifshitz scalar theory  the dispersion relation is  given by $\epsilon_k=\sqrt{\vec{k}^{2z}+m^{2z}}$ \cite{Alexandre:2011kr}.
By making use of Eqs. \eqref{fW} and \eqref{rho}, it is relatively straightforward to evaluate the Wightman power spectrum
\begin{eqnarray}\label{fWLif}
f^{W}(\omega)= \mathcal{N}(m, \beta, d, z)\frac{(\omega^2-m^{2z})^{\frac{d-1}{2 z}-1}}{z\sinh(\frac{\beta |\omega|}{2})}\Theta(|\omega|-m^z),
\end{eqnarray}
where the normalization factor $\mathcal{N}(m, \beta, d, z)$ can be determined by simply evaluating the following normalization condition
\begin{eqnarray}\label{norm}
\int\frac{d\omega}{2\pi}f^{W}(\omega)=1.
\end{eqnarray}

Let us also present a modified approach to find the results for a general class of theories which may lead to a great reduction in computing time in
numerical computations. First, combining eqs. \eqref{fW} and \eqref{rho} we have 
\begin{eqnarray}\label{fW1}
f^W(\omega)=\frac{\mathcal{N}}{\sinh \frac{\beta \omega}{2}}\int \frac{d^{d-1}k}{(2\pi)^{d-1}}\frac{1}{\epsilon_k}\left[\delta(\omega-\epsilon_k)-\delta(\omega+\epsilon_k)\right].
\end{eqnarray}
Inserting the above expression in eq. \eqref{norm} and changing the order of integration, the normalization factor then reads \footnote{Note that here without loss of generality, we set $\mathcal{N}\rightarrow   \mathcal{N}\Omega_{d-2}/\pi$.}
\begin{eqnarray}\label{norm1}
\mathcal{N}^{-1}=\int_0^{\infty} dk\frac{k^{d-2}}{\epsilon_k \sinh \frac{\beta \epsilon_k}{2}}.
\end{eqnarray}
Next, combining eqs. \eqref{mu2n} and \eqref{fW1} and following the similar steps one finds
 \begin{eqnarray}\label{mu2n1}
\mu_{2n}=\mathcal{N}\int_0^{\infty} dk\;k^{d-2}\frac{\epsilon_k^{2n-1}}{ \sinh \frac{\beta \epsilon_k}{2}}.
\end{eqnarray}
Finally inserting eq. \eqref{fW1} in the inverse Fourier transform of eq. \eqref{Ftrans}, the auto-correlation function becomes
\begin{eqnarray}\label{phi0}
\phi_{0}(t)=\mathcal{N}\int_0^{\infty} dk\;k^{d-2}\frac{\cos \epsilon_k t}{ \epsilon_k\sinh \frac{\beta \epsilon_k}{2}}.
\end{eqnarray}
In order to find K-complexity numerically we have to calculate the $n$th derivative of the above expression which can be simplified as follows
\begin{eqnarray}\label{phi0nth}
\frac{\partial^n	\phi_0(t)}{\partial t^n}=\mathcal{N}\int_{0}^{\infty}dk  \;k^{d-2}\;\epsilon_k^{n-1}\frac{\cos(\frac{n\pi}{2}+\epsilon_k t) }{\sinh\frac{\beta \epsilon_k}{2}},
\end{eqnarray}
which could be used to compute $\phi_n(t)$ by making use of Eq. \eqref{disSchrodinger}. It is then possible to compute the K-complexity and other physical quantities which we defined in the previous section as a function of time.

\subsection{Massless case}

\begin{figure}[h!]
	\begin{center}
		%\includegraphics[width=0.4\linewidth]{LB,m=0,z12}
		%\hspace*{1cm}
		\includegraphics[width=0.48\linewidth]{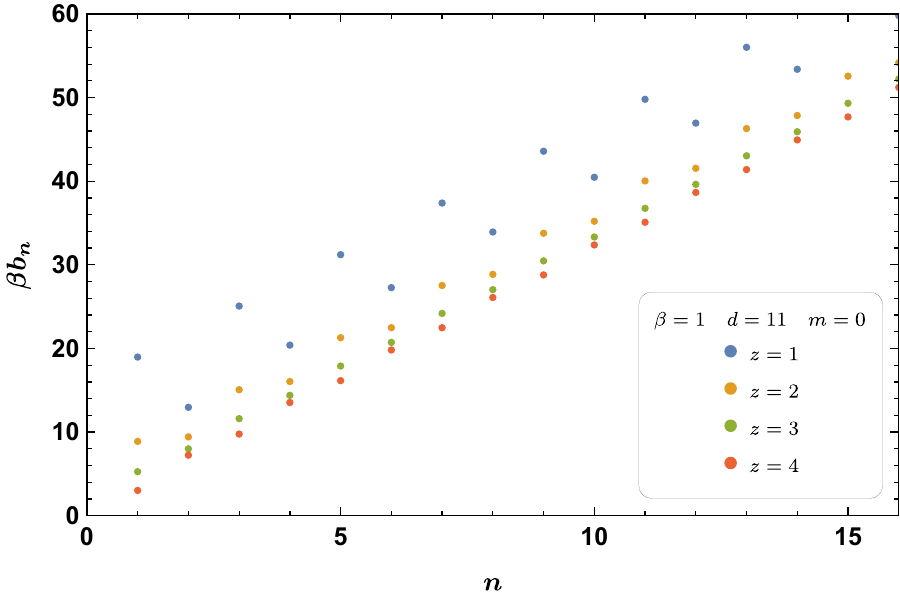}
  \hspace*{0.2cm}
		\includegraphics[width=0.48\linewidth]{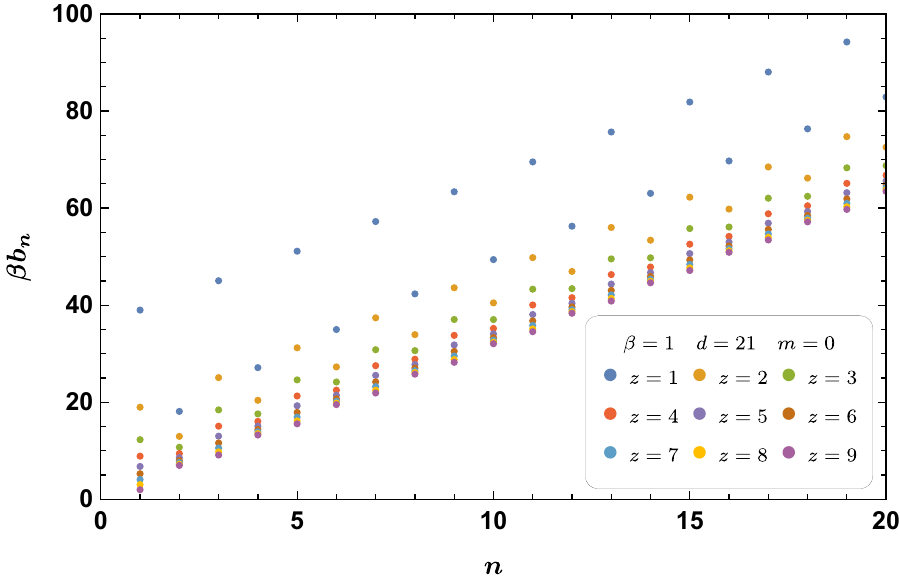}
  	\end{center}
	\caption{Lanczos coefficients in the massless regime for different values of $z$ and $d$. 
 %In order to present the results for several $z$ in one plot while maintaining the clarity and better resolution, the numerical results are given 
 %for $d=11$.
 As we see, although the slope is the same for all $z$, the
 $y$-intercept depends on $z$. In particular, as one increases $z$, the
  $y$-intercept decreases and the difference of $y$-intercepts for odd and even $n$ becomes less pronounced.}
	\label{fig:lcms}
\end{figure}

To explore the $z$-dependence of the Lanczos coefficients in our model, in what follows, we will consider the massless case, for which from Eq. \eqref{fWLif}, one gets 
\begin{eqnarray}\label{fwmassless}
f^{W}(\omega)= \frac{\pi  \beta}{(2^{\frac{d-1}{z}}-2)\Gamma\left(\frac{d-1}{z}-1\right)\zeta\left(\frac{d-1}{z}-1\right)}\frac{|\beta\omega|^{\frac{d-1}{z}-2}}{\sinh(\frac{\beta |\omega|}{2})}\Theta(|\omega|).
\end{eqnarray}
Here, we have used Eq. \eqref{norm} to fix the normalization factor ${\cal N}$. From this expression, one observes that in the high-frequency limit, the power spectrum becomes $f^{W}(\omega\rightarrow \infty) \sim e^{-\frac{\beta \omega}{2}}\omega^{\frac{d-1}{z}-2}$, which is, indeed, the scaling behavior we expect to have for an operator with dimension $2\Delta=\frac{d-1}{z}-1$ in a scale-invariant theory (for the CFT case see \cite{Dymarsky:2021bjq}). Plugging this expression into Eq. \eqref{mu2n}, the moments $\{\mu_{2n}\}$ are computed as follows
\begin{eqnarray}
\mu_{2n}=	\frac{\beta ^{-2n} \left(2\ 2^{1/z}-2^{\frac{d}{z}+2n}\right) \zeta \left(2n+\frac{d-z-1}{z}\right) \Gamma \left(2n+\frac{d-z-1}{z}\right)}{\left(2\ 2^{1/z}-2^{d/z}\right) \zeta \left(\frac{d-z-1}{z}\right) \Gamma \left(\frac{d-z-1}{z}\right)},
\end{eqnarray}
where $\zeta$ denotes the zeta function. Although the above expression for moments $\mu_{2n}$ looks very complicated, one can numerically evaluate the Lanczos coefficients using Eq. \eqref{recursion}. The results are depicted in figure \ref{fig:lcms} for several values of $z$ and $d$\footnote{
The scaling dimension of the scalar field is given by $\frac{d-z}{2}$. Restricting to positive scaling dimensions, studying larger values of $z$ is possible in $d>z$ dimensions. Meanwhile $d<z$ are well-behaving as long as an IR cut-off is introduced (see for instance \cite{MohammadiMozaffar:2017nri, He:2017wla, MohammadiMozaffar:2017chk, MohammadiMozaffar:2018vmk, MohammadiMozaffar:2019gpn, Hartmann:2021vrt, Mozaffar:2021nex, Mintchev:2022xqh, Mintchev:2022yuo}).
%Since in the equations we are considering, the number of the space-time dimensions and the critical exponent appears in the form of $d-z$, in order to study large $z$, usually, one needs to consider higher dimensions.
}. 

Interestingly enough, looking at the numerical results shown in figure \ref{fig:lcms}, one observes that the slope is independent of $d$ and $z$. Hence, non-relativistic scale invariance does not influence the rate of change of the Lanczos coefficients. Indeed, as far as the slope is concerned, one can see that for all cases, the best fit is given by $b_n=\frac{\pi}{\beta}n+\cdots$.

We note, however, that the $y$-intercept of different cases depends on $z$. More precisely, as one increases the critical exponent, the $y$-intercept decreases. Moreover, the staggering effect\footnote{The staggering effect
is defined through the fact that the Lanczos coefficients, $b_{n}$, are separated into two families for odd and even $n$ \cite{Yates:2020lin,Dymarsky:2021bjq,Bhattacharjee:2022vlt,Avdoshkin:2022xuw,Camargo:2022rnt}.}, which causes to have different $y$-intercepts for even and odd $n$'s for given cases, becomes less pronounced as we increase $z$. Actually, for large $n$, the best fit is
\begin{eqnarray}\label{LNLC}
b_{n}=\frac{\pi}{\beta}\left(n+\frac{d-2z-1}{2z}\right).
\end{eqnarray}
It is worth noting that considering the high-frequency limit of Eq. \eqref{fwmassless}, the above numerical fit is consistent with the prediction of \cite{Dymarsky:2021bjq} where it was shown that pole structure of $\phi_0(t)$ controls the asymptotic behavior of Lanczos coefficients.

\begin{figure}[h!]
	\begin{center}
		\includegraphics[width=0.48\linewidth]{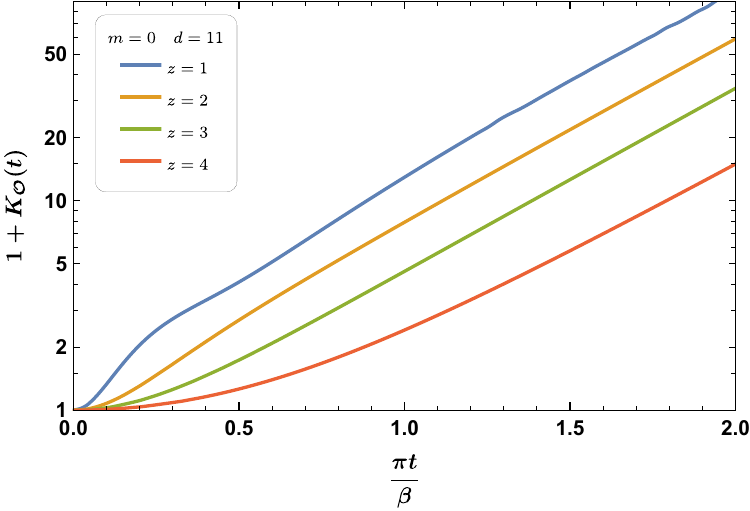}
		\end{center}
	\caption{Evolution of K-complexity in the massless regime for various values of the dynamical exponent. The complexity decreases as one increases $z$, though for 
 the late time, the slope is the same for all cases, which is given by $\frac{2\pi}{\beta}$.}
	\label{fig:KCm0}
\end{figure}

Let us now turn to the computation of the K-complexity in this setup using Eq. \eqref{KC}. To proceed, we note that from Eq. \eqref{fwmassless} and using inverse Fourier transformation, one can find $\phi_0(t)$ as follows
\begin{eqnarray}
\phi_0(t)=	\frac{\zeta \left(\frac{d-z-1}{z},\frac{i t}{\beta }+\frac{1}{2}\right)+\zeta \left(\frac{d-z-1}{z},\frac{1}{2}-\frac{i t}{\beta }\right)}{2 \left(2^{\frac{d-z-1}{z}}-1\right) \zeta \left(\frac{d-z-1}{z}\right)}.
\end{eqnarray}
It is also straightforward  to obtain a closed form for 
 $n$-th derivative of $\phi_0(t)$
\begin{eqnarray}
\frac{d^n\phi_0(t)}{dt^n}=	\frac{\Gamma \left(\frac{d-1-z}{z}+n\right) \left(\left(\frac{i}{\beta }\right)^n \zeta \left(\frac{d-1-z}{z}+n,\frac{i t}{\beta }+\frac{1}{2}\right)+\left(\frac{-i}{\beta }\right)^n \zeta \left(\frac{d-1-z}{z}+n,\frac{1}{2}-\frac{i t}{\beta }\right)\right)}{(-1)^{n+1} \left(2-2^{\frac{d-1}{z}}\right) \zeta \left(\frac{d-1}{z}-1\right) \Gamma \left(\frac{d-1}{z}-1\right)},
\end{eqnarray}
which could be used to compute probability amplitudes $\phi_n(t)$ numerically by making use of Eq. \eqref{disSchrodinger}. It is then possible to compute the different physical quantities we defined in the previous section. In particular, figure \ref{fig:KCm0} shows the K-complexity as a function of time in logarithmic scale for several values of $z$ with $d=11$. 
To produce these plots, we have used Eq. \eqref{KCapprox} with $n_{\rm max}=250$ to approximately compute K-complexity. 
 
From these numerical results, one observes that $K_{\mathcal{O}}(t)$ decreases with the dynamical exponent, which is perfectly consistent with the results illustrated in figure \ref{fig:lcms}. Moreover, since the vertical axis is in the logarithmic scale,  the linear growth corresponds to an exponential growth for the complexity. From Eq. \eqref{lambdaKbound}, representing our best fit for the Lanczos coefficients, one would expect that the slope of these curves is the same for all cases and is equal to $\frac{2\pi}{\beta}$. Indeed, our numerical results confirm our expectation for different values of $d$ and $z$.

% in table \ref{table:lambdaKmassless}support this behavior for several 
%\begin{table}[h]
%	\begin{center}
%		\begin{tabular}{|c|c|c|c|c|}
%			\hline
%			\diagbox[dir=SE,width=5 em]{$z$}{$d$}	& 8 &9  & 10 &11\\
%			\hline
%			1	&0.025  &0.018  &0.061 & 0.016\\
%			\hline
%			2	& 0.061 &0.034  &0.041&  0.024\\
%			\hline
%			3	& 0.003 & 0.062 &0.036& 0.034 \\
%			\hline
%		\end{tabular}
%		\caption{$\left(\lambda_K-\frac{2\pi}{\beta}\right)/\frac{2\pi}{\beta}$ %for different values of $d$ and $z$.}
%				\label{table:lambdaKmassless}
%	\end{center}
%
%\end{table}
%Further, let us note that for the examples in this table the linear fitting is %done in a finite time interval and hence the slope is always larger than %$\frac{2\pi}{\beta}$. 

\begin{figure}[h]
	\begin{center}
		\includegraphics[width=0.48\linewidth]{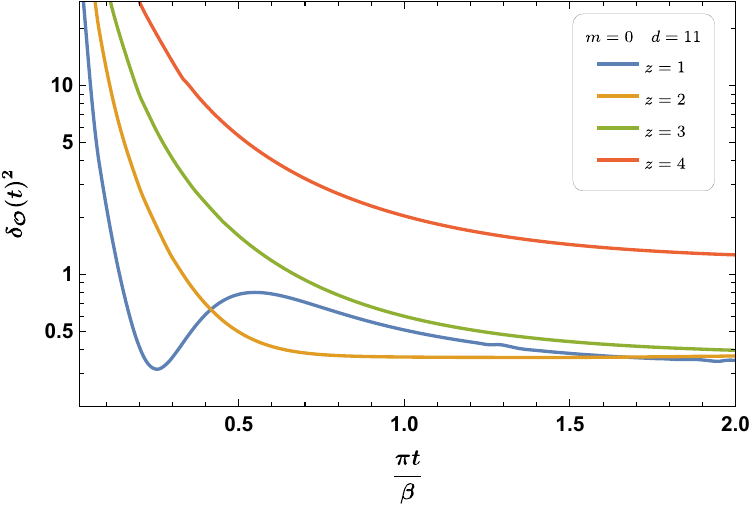}
\hspace*{0.2cm}
				\includegraphics[width=0.48\linewidth]{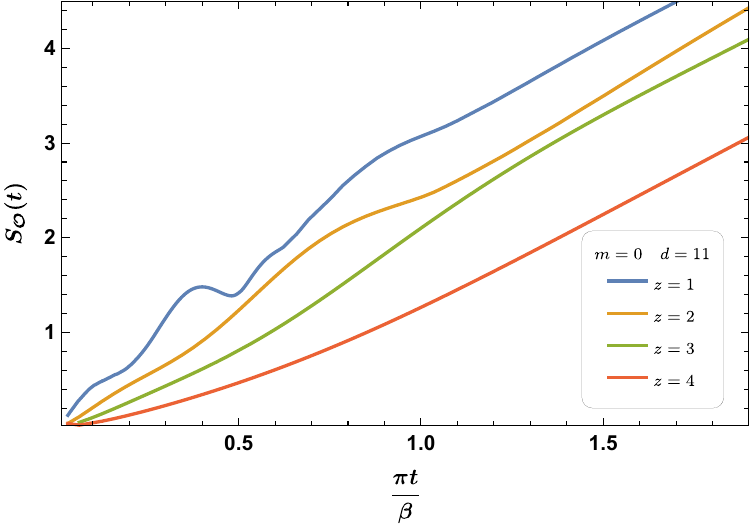}
				
		\end{center}
\caption{K-variance (left) and K-entropy (right) as a function of time in the massless regime for various values of the dynamical exponent. }
\label{fig:KVKEm0}
\end{figure}

To close this subsection, we present our numerical results for  K-variance and K-entropy, defined in equations \eqref{Kvariance} and \eqref{Kentropy}, respectively,  in figure \ref{fig:KVKEm0}. Note that all of the curves for $\delta_{\mathcal{O}}(t)$ stabilize to a constant value at late times, and further, the fluctuations become less pronounced for larger values of the dynamical exponent. As for K-entropy, one gets late time linear growth with the same slope for different $z$. Indeed, our numerical results show that the slope is again given by $\frac{2\pi}{\beta}$. 

% Clearly, in both cases the fluctuations stabilize to a constant value at late times and suppress for $z>1$.

\subsection{Massive case}\label{largemass}
\begin{figure}[h!]
	\begin{center}
						%\includegraphics[width=0.47
      %\linewidth]{LB,z=2,Perturb}
		%\hspace*{0.4cm}
		\includegraphics[width=0.54\linewidth]{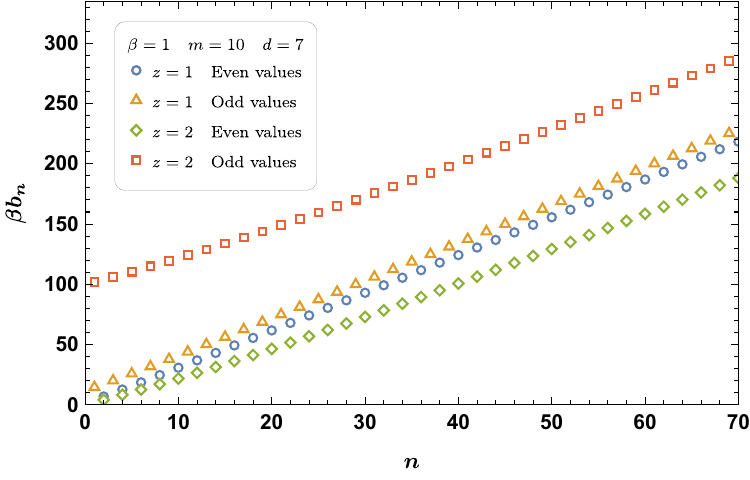}
\end{center}
\caption{Lanczos coefficients in the large mass regime. In this case the $y$-intercept is affected by non-zero mass.}
\label{fig:mbetaOE}
\end{figure}

In this section, we consider a massive scalar field to explore how the non-zero mass could affect the results presented in the previous section. To highlight these effects, in what follows we will consider large mass limit in low temperature regime. To be more concrete, we will consider the case in which $\beta m^z\gg 1$ and hence, from Eq. \eqref{fWLif}, the normalized Wightman power spectrum reads
%eqs. \eqref{fW} and \eqref{rho} we have $\sinh(\frac{\beta |\omega|}{2})\sim
%\frac12  e^{\frac{\beta |\omega|}{2}}$. Then following Eq.  
\begin{eqnarray}\label{fWlargemass}
f^{W}(\omega)=\frac{z\pi^{3/2}\beta^{\frac{d-z-1}{2z}}}{2^{\frac{d-z-1}{z}}m^{\frac{d-z-1}{2}}K_{\frac{d-z-1}{2z}}\left(\frac{\beta m^z}{2}\right)\Gamma\left(\frac{d-1}{2z}\right)}\frac{(\omega^2-m^{2z})^{\frac{d-1}{2 z}-1}}{z}e^{-\frac{\beta |\omega|}{2}}\Theta(|\omega|-m^z).
\end{eqnarray}
It is found useful to express our results in terms of the dimensionless parameter $s=\beta m^z$. Plugging the above expression into Eq. \eqref{mu2n}, one arrives at
\begin{eqnarray}
\mu_{2n}=&&\frac{\sqrt{\pi}\left(\frac{s}{4}\right)^{\frac{d+z-1}{2z}}}{K_{\frac{d-z-1}{2z}}\left(\frac{s}{2}\right)}\left(\frac{2}{\beta}\right)^{2n}\Bigg(\frac{\Gamma\left(\frac{1+z-d}{2z}-n\right)}{\Gamma\left(\frac{1}{2}-n\right)}\left(\frac{s}{2}\right)^{2n-1}{}_pF_q\left(n+\frac{1}{2}; \frac{1}{2}, \frac{d-1}{2z}+n+\frac{1}{2}; \left(\frac{s}{4}\right)^2\right)\nonumber\\
&&+2\frac{\Gamma\left(\frac{d-1}{z}+2n-1\right)}{\Gamma\left(\frac{d-1}{2z}\right)}\left(\frac{s}{2}\right)^{\frac{1-d}{z}}{}_pF_q\left(1+\frac{1-d}{2z}; 1-n+\frac{1-d}{2z}, \frac{1-d}{2z}+\frac{3}{2}-n; \left(\frac{s}{4}\right)^2\right)\nonumber\\
&&-\frac{\Gamma\left(\frac{1-d}{2z}-n\right)}{\Gamma\left(-n\right)}\left(\frac{s}{2}\right)^{2n}{}_pF_q\left(n+1; \frac{3}{2}, \frac{d-1}{2z}+n+1; \left(\frac{s}{4}\right)^2\right)\Bigg),
\end{eqnarray}
where ${}_pF_q$ is the regularized generalized hypergeometric function. This expression can be used to obtain the Lanczos coefficients numerically from Eq. \eqref{recursion}. The numerical results are shown in figure \ref{fig:mbetaOE}.
As it is evident from this figure, the Lanczos coefficients for the massive case exhibit qualitatively similar behavior as that of the massless one with two interesting features. First, we note that, similar to the massless case, the slope of the curves for odd and even $n$ are the same. Actually, in the present case, the best fit is given by 
\begin{eqnarray}\label{bnfitlargem}
b_{n}=\;\bigg\{\hspace{-0.5cm}\begin{array}{rcl}
&\alpha_{{\rm o}}\;n+\gamma_{{\rm o}} &\,\,\, {\rm odd}\; n,\cr
%&\cr
&\alpha_{{\rm e}}\;n+\gamma_{{\rm e}} &\,\,\, {\rm even}\; n,
\end{array}
\end{eqnarray}
where based on our numerics we have $\alpha_{{\rm e}}=\alpha_{{\rm o}}\equiv \alpha\sim \frac{\pi}{\beta}$.\footnote{We would like to thank the anonymous referee for a useful comment on this point.}
 Moreover, $\alpha$ is $z$ independent and in particular the "staggering" behavior is also observed in the non-relativistic case. 

The second new feature we find for the massive case is that, unlike the massless case, the difference of $y$-intercepts for even and odd curves increases as one increases the critical exponent, indicating that for non-zero mass the difference $|\gamma_{{\rm o}}-\gamma_{{\rm e}}|$ is of the order of $m^z$. This observation may be confirmed by evaluating $b_n$s in a series expansion for $s \gg n$. To be concrete for $z=2$ and $d=5$ in this limit, one finds 
\begin{eqnarray}\label{bnapproxz2d5}
b_{n}=m^2\;\Bigg\{\hspace{-0.5cm} \begin{array}{rcl}
&1+\frac{2n}{s}+\frac{2n^2}{s^2}-\frac{2n(n^2+1)}{s^3}+
\cdots &\,\,\,n\in 2k+1,\\
&\frac{2n}{s}+\frac{2n^2}{s^2}-\frac{2n^3}{s^3}+\cdots 
&\,\,\,n\in 2k,
\end{array}
\end{eqnarray}
resulting in $|\gamma_{{\rm o}}-\gamma_{{\rm e}}|
\approx m^2$.

\begin{figure}[h!]
	\begin{center}
			\includegraphics[width=0.4\linewidth]{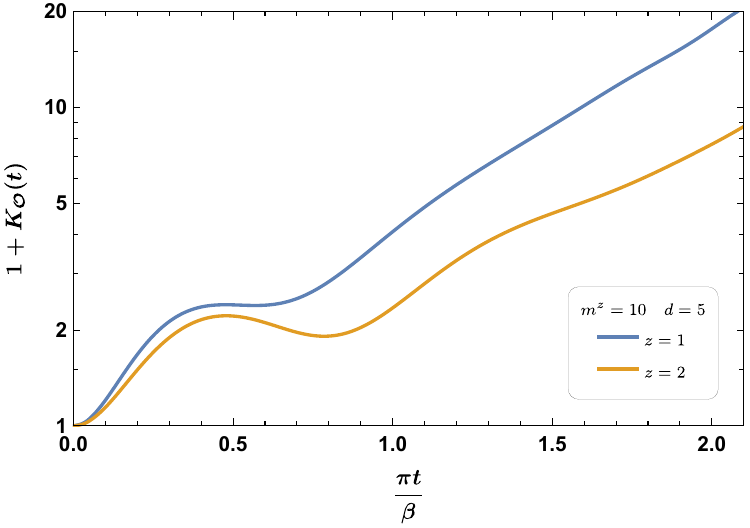}
					\hspace*{.01cm}
			\includegraphics[width=0.4\linewidth]{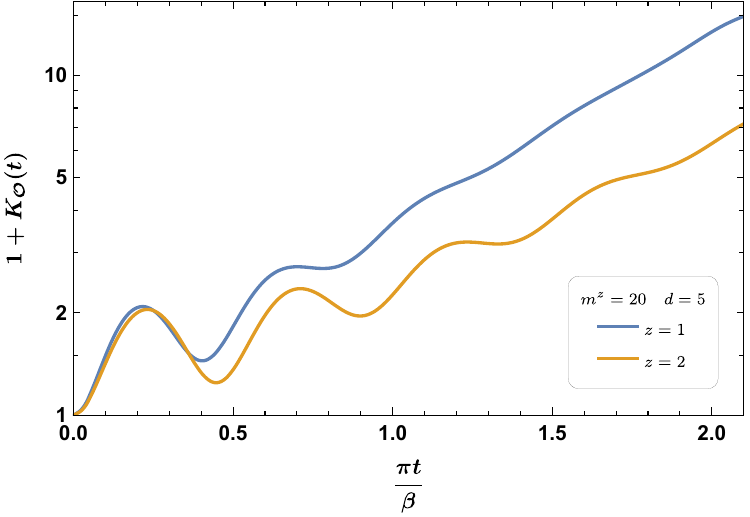}
				%	\hspace*{.01cm}
		%\includegraphics[width=0.32\linewidth]{KCz=1,2m^z=30d=5}
	\end{center}
	\caption{K-complexity in the large mass limit for different values of $m$ and $z$ with $d=5$. Here we set $\beta=1$. At early times the results are independent of the critical 
 exponent, though at late times it has significant effects.}
	\label{fig:KCm10m50m100}
\end{figure}

\begin{figure}[h!]
	\begin{center}
		\includegraphics[width=0.4\linewidth]{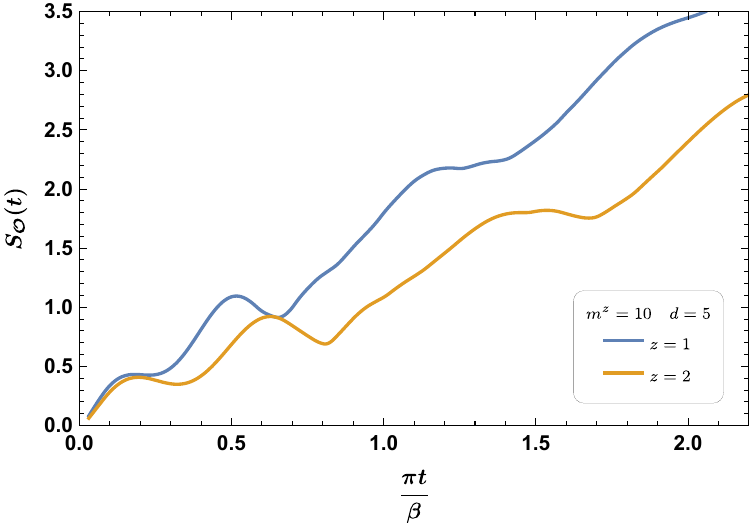}
		\hspace*{.01cm}
		\includegraphics[width=0.4\linewidth]{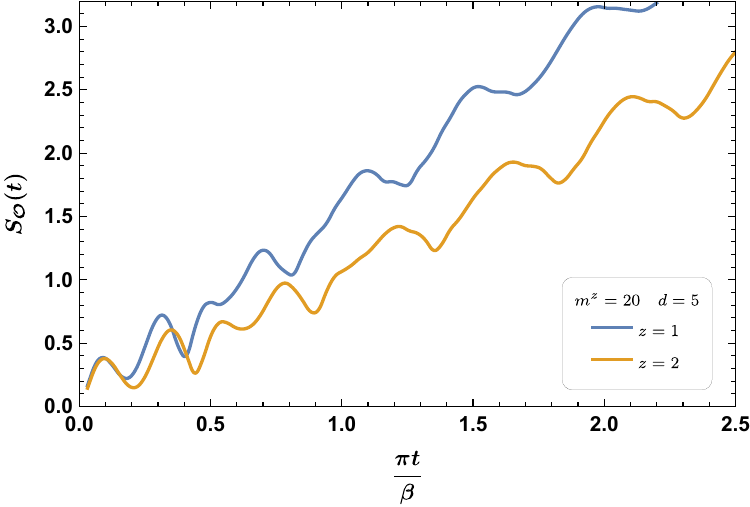}
%		\hspace*{.01cm}
	%	\includegraphics[width=0.32\linewidth]{KSz=1,2m^z=30d=5}
	\end{center}
	\caption{Evolution of K-entropy for several values of the mass and dynamical exponent.}
	\label{fig:KEntropymassive}
\end{figure}

Let us now compute K-complexity for the massive case. To do so, one may compute the auto-correlation function from Eq. \eqref{fWlargemass} as follows
\begin{eqnarray}\label{phi0}
\phi_0(t)=\frac{\left(1-\frac{2it}{\beta}\right)^{\frac{1-d+z}{2z}}K_{\frac{1-d+z}{2z}}\left(m^z\frac{\beta-2it}{2}\right)+\left(1+\frac{2it}{\beta}\right)^{\frac{1-d+z}{2z}}K_{\frac{1-d+z}{2z}}\left(m^z\frac{\beta+2it}{2}\right)}{2K_{\frac{1-d+z}{2z}}\left(\frac{\beta m^z}{2}\right)},
\end{eqnarray}
where $K_n(x)$ is the Bessel function of the second kind. It is then straightforward to compute K-complexity numerically in the large mass limit by making use of Eqs. \eqref{disSchrodinger} and \eqref{KC}. To perform our computations, we have used Eq. \eqref{KCapprox} with $n_{\rm max}=100$, which is a good approximation for the time interval we have considered. The corresponding numerical results for different $m$ and $z$ are shown in figure \ref{fig:KCm10m50m100}. Similarly, one could also compute K-entropy numerically, and  the corresponding result is depicted in figure \ref{fig:KEntropymassive}.
%and \ref{fig:KVariancemassive}, respectively. 
%\begin{figure}[h]
%	\begin{center}
%		\includegraphics[width=0.4\linewidth]{Kvarz=1,2m^z=10d=5}
%		\hspace*{.01cm}
%		\includegraphics[width=0.4\linewidth]{Kvarz=1,2m^z=20d=5}
%		%\hspace*{.01cm}
%		%\includegraphics[width=0.32\linewidth]{Kvarz=1,2m^z=30d=5}
%	\end{center}
%	\caption{Evolution of K-variance for several values of the mass %and dynamical exponent.}
%	\label{fig:KVariancemassive}
%\end{figure}

From our numerical results shown in these figures, one observes that the general behavior for K-complexity and K-entropy are the same as those in the massless case. Namely, while complexity
has exponential growth, the K-entropy has linear growth at late times. We note, however, at early times, there is an oscillatory 
behavior that is associated with non-zero mass. Actually,
these oscillations are originating from the oscillatory behavior of 
probability amplitudes $\phi_n(t)$ \cite{Camargo:2022rnt}. To see
this point better, it is useful to write the explicit form of the 
auto-correlation function Eq. \eqref{phi0} for $z=2$ 
\begin{eqnarray}\label{phi0massivez2}
\phi_0(t)=\frac{\beta^2\cos (m^2t)-2\beta t\sin (m^2t)}{\beta^2+4t^2}.
\end{eqnarray}

\begin{figure}[h!]
	\begin{center}
		\includegraphics[width=0.49\linewidth]{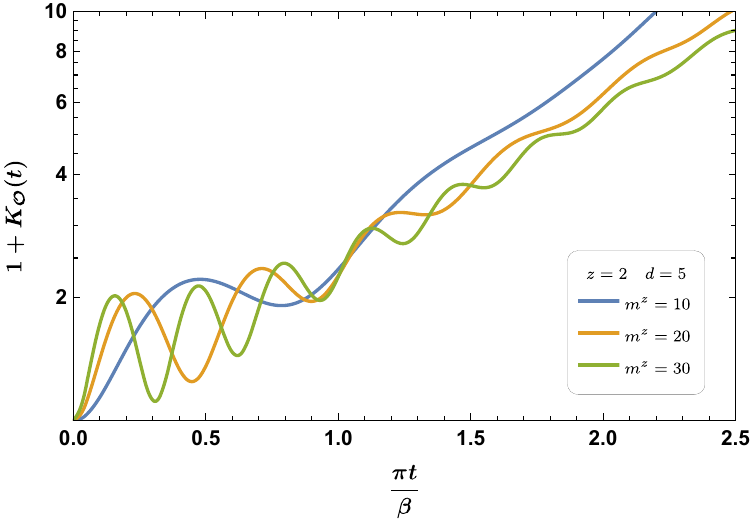}
	\end{center}
	\caption{Evolution of K-complexity for several values of the mass with $z=2$.}
	\label{fig:Kcomplexityz2mass}
\end{figure}

It is also worth noting that while at early times the results are independent of $z$, at late times the critical exponent has significant effects such that as one increases $z$, both complexity and entropy decrease. This is due to the fact that the slope of the curves is affected by the non-zero mass so that it is always smaller than that of the massless case, moreover, from dimensional analysis one finds that  the mass dependence is in the form of $m^z$. Note also that the period of the oscillations in the oscillatory region is given by $m^{-z}$ so that the amplitude of oscillation becomes less pronounced at later times. Let us examine in more detail the mass-dependence of K-complexity as shown in figure \ref{fig:Kcomplexityz2mass}. We see that by increasing the mass parameter, the K-complexity decreases. For the relativistic case with $z=1$, this is intuitive in the sense that the correlation length of a massive field is decreased with the mass parameter and so is its complexity. Similarly, for the larger values of the dynamical exponent, we  expect that the effective length of quantum fluctuations is decreased by mass, and hence the K-complexity decreases. 
Although we have not presented the results for K-variance, we have numerically computed it and its general behavior exhibits similar features to those discussed for complexity and entropy.

%\begin{figure}[h]
%	\begin{center}
%		\includegraphics[width=0.49\linewidth]
%{KCz1m10,20,30}
%		\hspace*{.01cm}
%		\includegraphics[width=0.49\linewidth]
%{KCz2m10,20,30}
%	\end{center}
%	\caption{K-complexity for $z=1$ (left) and $z=2$ 
%(right) with $d=5$. Results are shown for several %values of the mass parameter.}
%	\label{fig:KCz1z2}
%\end{figure}

\section{Lanczos coefficients and Krylov complexity in the presence of UV cutoff}\label{UVCUT}

In this section, we will study the effect of having a dimensionful scale in the model on the behavior of Lanczos coefficients and, consequently, K-complexity. There are several ways one may have a non-trivial dimensionful scale. In the previous section, we have already studied  the case where the dimensionful parameter is given by non-zero mass, which acts as an IR cutoff, and its effects on the slopes of the linear growth of Lanczos coefficients at large values of $n$ were observed.

We note that there are several other ways to have a new scale in the model. For example, we can achieve this by putting the model in a compact space or discretizing the model by putting it on a lattice. In these cases, the scale is given by the curvature radius of the compact space or lattice spacing, respectively. Another way to have a non-trivial scale is to add a hard or soft cutoff to the theory.

In all cases, we expect that the behavior of Lanczos coefficients will be influenced by the dimensionful scale, the nature of which may depend on the way the scale is added to the model.
In what follows, we will study the effect of having a non-zero hard cutoff and the case where the model is put on a lattice.

\subsection{Krylov complexity with hard UV cutoff}

\begin{figure}[h!]
	\begin{center}
		\includegraphics[width=0.49\linewidth]{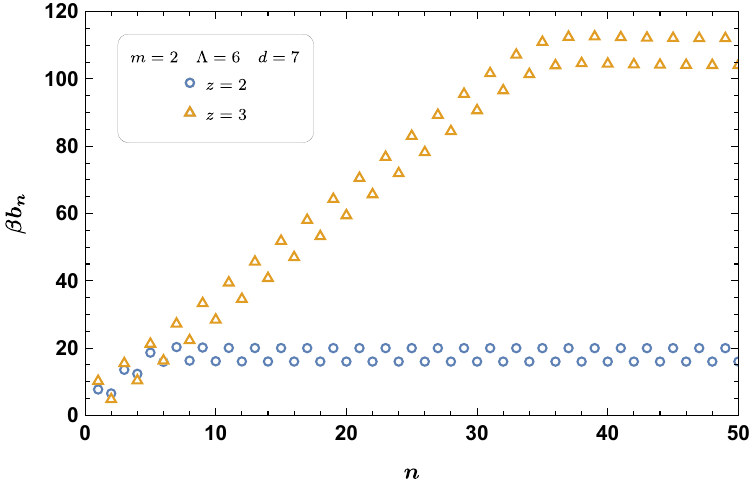}
		\hspace*{.01cm}
		\includegraphics[width=0.49\linewidth]{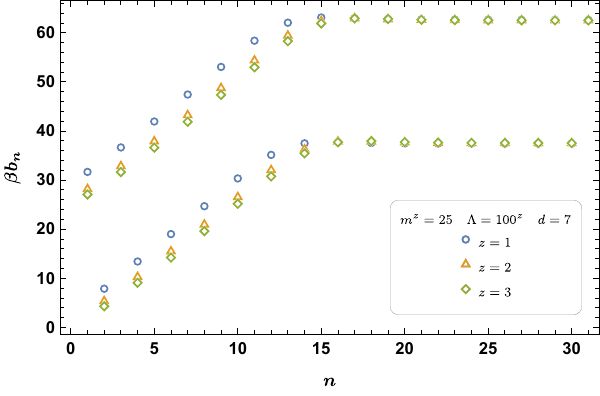}
	\end{center}
	\caption{Lanczos coefficients in $d=5$ for different values of $z$, $m$ and $\Lambda$.}
	\label{fig:UVLC}
\end{figure}

In this section, we will examine the effects of a finite UV cutoff in continuous momentum space on the behaviors of Lanczos coefficients and K-complexity. In particular, we will consider a  UV cutoff on the integral upper bound in Eq. \eqref{fW}. Hence, the Wightman power spectrum will undergo the following change
\begin{eqnarray}
f^W(\omega)=\frac{1}{\sinh \frac{\beta \omega}{2}}\int_{0}^{\Lambda^z} \frac{d^{d-1}k}{(2\pi)^{d-1}}\rho(\omega, \vec{k})=
\mathcal{N}\frac{(\omega^2-m^{2z})^{\frac{d-2z-1}{2z}}}{z|\sinh(\frac{\beta \omega}{2})|}\Theta(|\omega|-m^z)\Theta(\Lambda^z-|\omega|).
\end{eqnarray}
In the limit of $1 \ll \beta m^z\ll \beta \Lambda^z$, the above equation reads
\begin{eqnarray}
f^{W}(\omega)=\mathcal{N}(m,\beta,d)
(\omega^2-m^{2z})^{\frac{d-1}{2 z}-1} e^{-\frac{\beta |\omega|}{2}} \Theta(|\omega|-m^z)\Theta(\Lambda^z-|\omega|)\,.
\end{eqnarray}
By making use of Eq. \eqref{mu2n}
it is straightforward to compute 
the moments $\{\mu_{2n}\}$. In particular, for  $z=2$ and $d=5$, one finds
\begin{eqnarray}
\mu_{2n}= 2^{2 n} \beta ^{-2 n}\frac{ \Gamma \left(2 n+1,\frac{m^2 \beta }{2}\right)-\Gamma \left(2 n+1,\frac{\beta  \Lambda ^2}{2}\right)}{e^{-\frac{1}{2} \left(\beta  m^2\right)}-e^{-\frac{1}{2} \left(\beta  \Lambda ^2\right)}}.
\end{eqnarray}
Although it is impossible to find an analytic 
expression for Lanczos coefficients for a general choice of the parameters, one can use 
the recursion relation Eq. \eqref{recursion} to find them numerically. The corresponding numerical results for different values of $z$ with the finite $\Lambda$ are shown in figure \ref{fig:UVLC}.
It is clear from this figure that the presence of a UV cutoff significantly modifies the behavior of Lanczos coefficients. The situation appears to be quite similar to that of a relativistic case 
\cite{{Avdoshkin:2022xuw},{Camargo:2022rnt}}. Specifically, the system undergoes a phase of linear growth as described by Eq. \eqref{bnfitlargem}, but eventually saturates to a constant value as $n$ becomes large. This saturation value is proportional to the UV cutoff.
More precisely, the saturation value can be expressed as $b_{\rm s}^{}\approx \frac{\Lambda^z\pm m^z}{2}$, where the sign should be chosen as $+(-)$ for odd (even) values of $n$. 
In addition, the transition occurs at a sharp saturation point $n=n_{\rm s}$, which can be estimated as
%can be estimated as follows: Assuming to have the linea 
%growth all the way to the saturation point, $n_{\rm s}$ is %determined by the following equation
%\begin{eqnarray}
%\alpha_{{\rm o, e}}\;n_s+\gamma_{{\rm o, e}}=\frac{\Lambda^z\pm m^z}{2}.
%\end{eqnarray}
%Note that $\alpha_{{\rm o}}=\alpha_{{\rm e}}\equiv \alpha$ and hence solving the above equation we find
\footnote{Here we use the fact that $\alpha_{{\rm o}}\approx \alpha_{{\rm e}}\equiv \alpha$ and $\gamma_{\rm e, o} \propto m^z\ll \Lambda^z$.} 
\begin{eqnarray}
n_s=\frac{\Lambda^z-\gamma_{{\rm o}}-\gamma_{{\rm e}}}{2\alpha}\approx\frac{\Lambda^z}{2\alpha},
\end{eqnarray}
which perfectly matches the numerical results. We see that in this case for larger values of the dynamical exponent both $b_s$ and $n_s$ increase.

In the right panel of figure \ref{fig:UVLC}, we plot Lanczos coefficients using a different choice of the mass and UV cutoff, which depends on the dynamical exponent, i.e., $m^z$ and $\Lambda^z$. It can be observed that the Lanczos coefficients exhibit linear growth with increasing $n$, followed by saturation to a constant value. 
Furthermore, at a given value of $n$, both the growth rate and the saturation value appear to be approximately independent of the critical exponent. In fact, their dependence on $z$ is implicitly determined by the choice of mass and UV scale.
%at a fixed value $b_{\rm n}=b_{\rm s}$ at a sharp saturation point $n=n_{\rm s}$. Using this new choice we see that both $b_{\rm s}$ and $n_{\rm s}$ are universal and independent of the dynamical exponent. 

\begin{figure}[h]
	\begin{center}
		\includegraphics[width=0.49\linewidth]{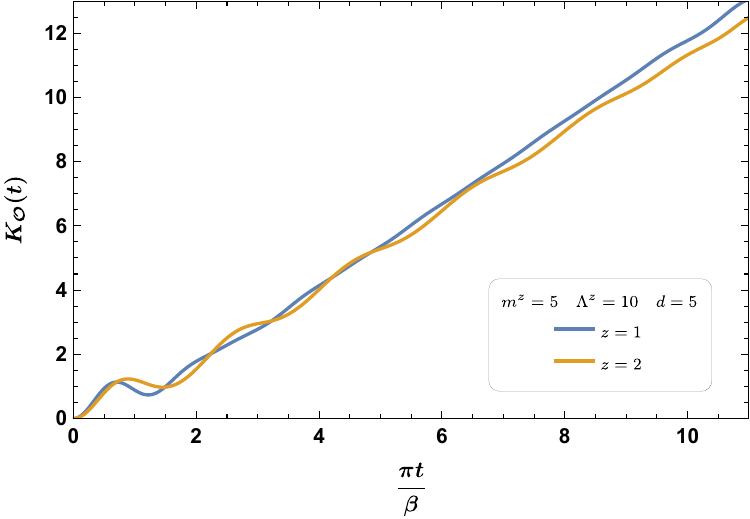}
		\hspace*{.01cm}
		\includegraphics[width=0.49\linewidth]{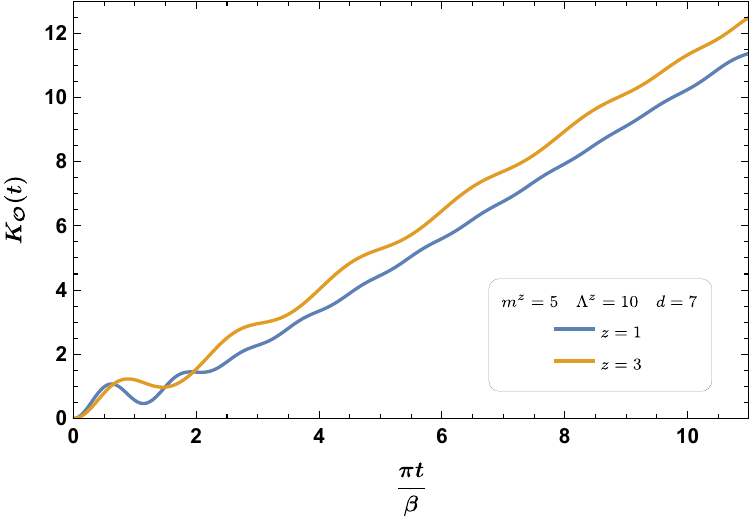}
	\end{center}
	\caption{K-complexity in the presence of a hard UV cutoff for different values of $z$ with $\beta=1$. In this case, the complexity exhibits exponential growth followed by linear growth at late times.}
	\label{fig:KCm50m100}
\end{figure}

Let us now consider the K-complexity. The procedure  is the same as what was done in the previous section. The numerical results are depicted in Figure \ref{fig:KCm50m100}.
As expected, the behavior of the complexity exhibits exponential growth at early times, followed by linear growth at later times. These phases are associated with linear and saturation phases of Lanczos coefficients, respectively. Our numerical results make it clear that for $z>1$ the oscillatory behavior of K-complexity persist for a longer time comparing with that for $z=1$ case.

%which is based on evaluating  Next, we turn our attention to %investigate how K-complexity influenced by the presence of %finite UV cutoff. From eqs. \eqref{phi0} and %\eqref{fwmassless}, we can find $\phi_0(t)$, e.g., for $z=2$ %and $d=5$ as follows
%\begin{eqnarray}
%\phi_0(t)=\frac{\beta  e^{\frac{\beta  m^2}{2}} \left(\beta  %\cos \left(\Lambda ^2 t\right)-2 t \sin \left(\Lambda ^2 %t\right)\right)-\beta  e^{\frac{\beta  \Lambda ^2}{2}} %\left(\beta  \cos \left(m^2 t\right)-2 t \sin \left(m^2 %t\right)\right)}{\left(\beta ^2+4 t^2\right) %\left(e^{\frac{\beta  m^2}{2}}-e^{\frac{\beta  \Lambda %^2}{2}}\right)},
%\end{eqnarray}
%which in the limit $\lambda \rightarrow\infty$ reduces to Eq. %\eqref{phi0massivez2} as expected. We then combine the above %expression with eqs. \eqref{disSchrodinger} and \eqref{KC} to %compute the K-complexity numerically. Figure %\ref{fig:KCm50m100} presents a plot of $K_{\mathcal{O}}(t)$ %for several values of the dynamical exponent with $\beta=1$. %We see that K-complexity grows linearly rather than %exponentially as expected. Indeed, as discussed in %\cite{Parker:2019} the linear growth of $K_{\mathcal{O}}(t)$ %is due to the saturation behavior of the Lanczos coefficients.

\begin{figure}[h]
	\begin{center}
		\includegraphics[width=0.49\linewidth]{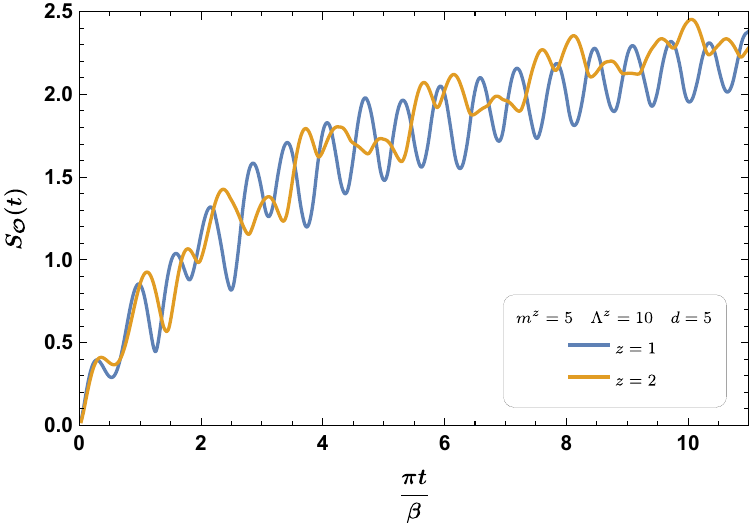}
		\hspace*{.01cm}
	\end{center}
	\caption{K-entropy in the presence of a hard UV cutoff for different values of $z$ with $\beta=1$.}
	\label{fig:kecut}
\end{figure}

It is also interesting to compute the K-entropy in the presence of a hard UV cutoff, as shown in figure \ref{fig:kecut} for different values of $z$.
%we plot the K-variance and K-entropy in the presence of a hard %UV cutoff for different values of $z$ and $d$. Again, we see %that $\delta_{\mathcal{O}}(t)$ stabilizes to a constant value %at late times.
%\begin{figure}[h]
%	\begin{center}
%		\includegraphics[width=0.49\linewidth]{KvarHCd5z12}
%		\hspace*{.01cm}
%		\includegraphics[width=0.49\linewidth]{KvarHCd7z13}
%	\end{center}
%	\caption{Evolution of K-variance in the presence of a hard %UV cutoff for different values of $z$ with $\beta=1$.}
%	\label{fig:kvarcut}
%\end{figure}
It exhibits an oscillatory behavior, which is a consequence of the non-zero mass, and approximately follows a logarithmic scaling due to the presence of a hard UV cutoff. This particular behavior is related to the linear growth of complexity at later times and will be discussed further in section \ref{Conclusions}.

\subsection{Krylov complexity for the Lifshitz harmonic model}\label{LCinLHM}

\begin{figure}[h]
	\begin{center}
		\includegraphics[width=0.49\linewidth]{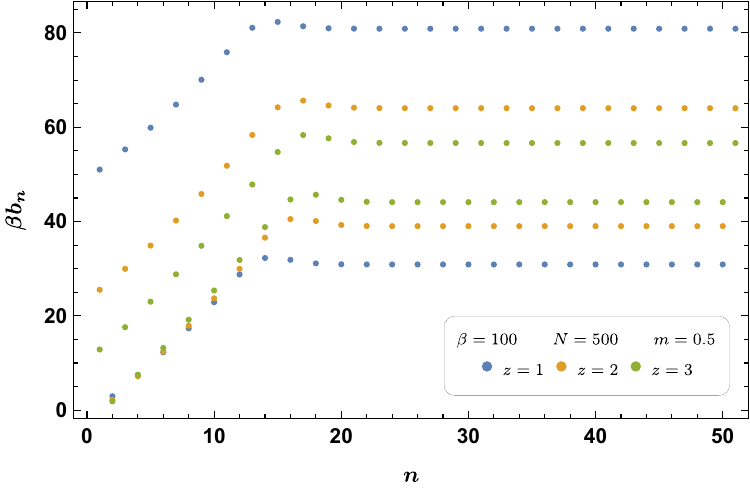}
		\hspace*{.1cm}
		\includegraphics[width=0.49\linewidth]{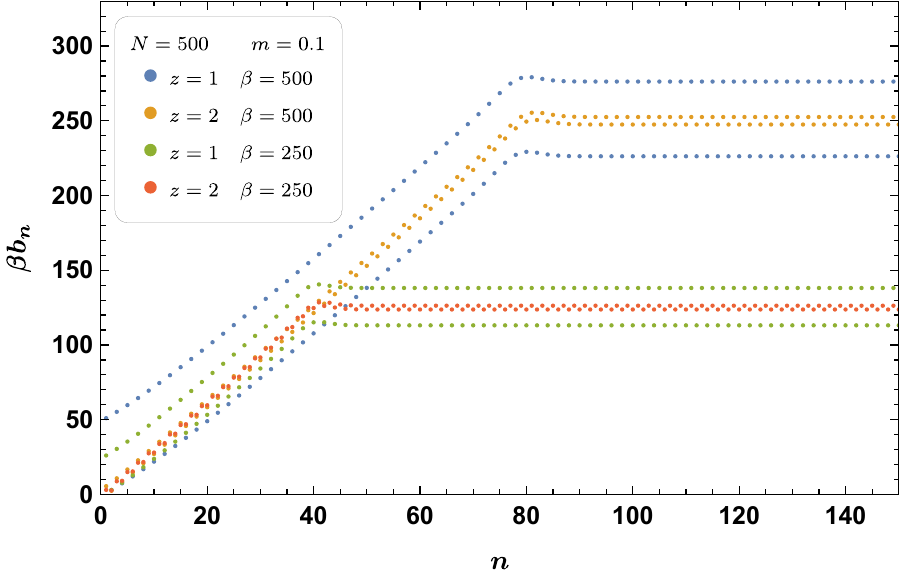}
	\end{center}
	\caption{Lanczos coefficients in Lifshitz harmonic model for different values of the parameters.}
	\label{fig:LClattice1}
\end{figure}

In this section, we will study a discretized version of our model with a finite lattice spacing $a$, which results in a UV cutoff of the form $\Lambda \sim \frac{1}{a}$. To be specific, we will consider a one-dimensional lattice with periodic boundary conditions. As mentioned earlier, in this case, the corresponding dispersion relation is given by Eq. \eqref{Lifdispersionlattice}, and thus the Wightman power spectrum takes the form\begin{eqnarray}\label{fwlattice}
f^{W}(\omega)=\mathcal{N}\sum_{k=1}^N\frac{1}{\sinh(\frac{\beta \omega}{2})}\frac{1}{\epsilon_k}[\delta(\omega-\epsilon_k)-\delta(\omega+\epsilon_k)],\qquad {\rm with}\qquad \epsilon_k=\sqrt{\sin(\frac{\pi k}{N})^{2z}+m^{2z}}.
\end{eqnarray}
Using Eq. \eqref{mu2n}, one may compute  the moments $\{\mu_{2n}\}$ on lattice as follows
\be
	%&= \int_{-\infty}^{\infty}d\omega \omega^{2n} f^{W}(\omega)\\
	\mu_{2n}=\frac{\mathcal{N}}{2\pi}\sum_{k=1}^N\int_{-\infty}^{\infty}d\omega  \frac{\omega^{2n}}{\epsilon_k \sinh(\frac{\beta \omega}{2})}[\delta(\omega-\epsilon_k)-\delta(\omega+\epsilon_k)]
	=\frac{\mathcal{N}}{\pi}\sum_{k=1}^N  \frac{\epsilon_k^{2n-1}}{ \sinh(\frac{\beta \epsilon_k}{2})}
\ee 
By substituting the above expression into Eq. \eqref{recursion}, one can numerically compute the Lanczos coefficients for various values of the parameters of the model. The corresponding results for different values of $\beta, m, z$ and $N$ are depicted in figure \ref{fig:LClattice1}. 
By making use of these  numerical results, several observations may 
be made. 
First, the same as that of hard UV cutoff, one 
observes that the Lanczos coefficients grow linearly with $n$ and then saturate to a constant value, \textit{i.e.}, $b_s$, for large $n$. Assuming that the saturation occurs at $n_s$, one finds that both $b_s$ and $n_s$ are decreasing functions of the temperature. Based on our results, this dependence is linear and the slope of the curve of $b_s$ or $n_s$ as a
function of $\beta$ is independent of the dynamical exponent.  
 Moreover, during the linear growth phase, the slope (approximately) remains constant, independent of the parameters $\beta$ and $z$. Interestingly, a staggering effect is observed, which decreases with increasing $z$.
%Note that in these plots we consider $m<1$, so comparing to section \ref{largemass} we expect that the separation of Lanczos coefficients between odd and even $n$ is proportional to $m^z$. Finally, in the right panel of figure \ref{fig:LClattice2}, we plot Lanczos coefficients for various $z$ with a varying mass parameter, i.e., $m^z$. In this case we see that all curves coincide and show a universal behavior.
In the linear phase, a linear fitting for $b_n$ can be proposed, as in Eq. \eqref{bnfitlargem} and again the slopes are independent of mass. 

\begin{figure}[h]
	\begin{center}
		\includegraphics[width=0.45\linewidth]{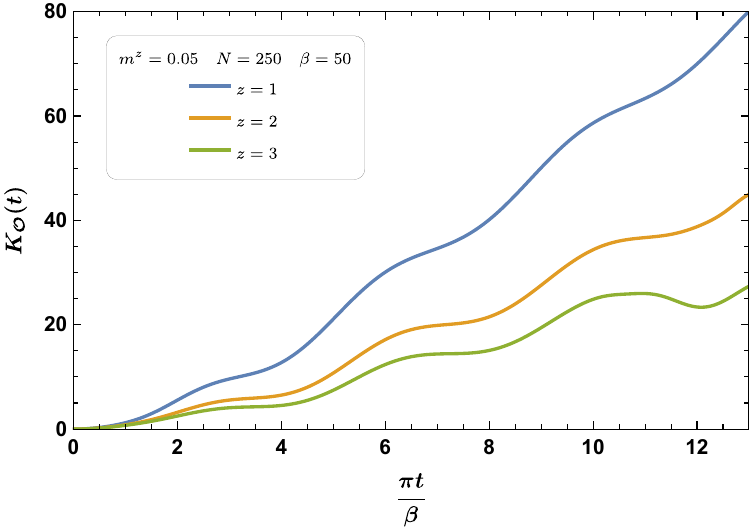}
		\hspace*{.01cm}
		\includegraphics[width=0.45\linewidth]{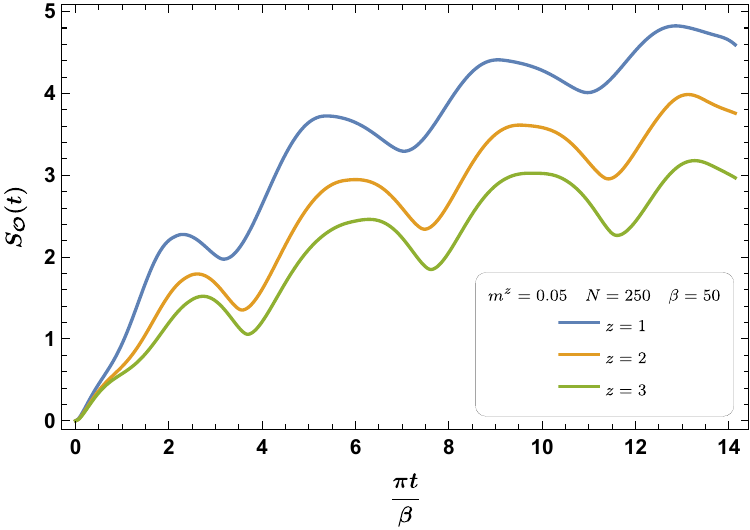}
		\hspace*{.01cm}
	\end{center}
	\caption{ K-complexity (Left) and K-entropy (right) 
 for the  Lifshitz harmonic model.}
	\label{fig:lattice measures}
\end{figure}

Let us now turn our attention to the computation of the K-complexity and K-entropy in this setup, using Eqs. \eqref{KC} and \eqref{Kentropy} . To proceed, we note that from Eq. \eqref{fwlattice}, one can find $\phi_0(t)$ as follows
\begin{align}
\phi_0(t)=\frac{\mathcal{N}}{\pi}\sum_{k=1}^N  \frac{ 1}{\epsilon_k }\csch\left(\frac{\beta  \epsilon_k }{2}\right) \cos (\epsilon_k \;t ).
\end{align}
It is then straightforward to compute different measures in the Krylov basis using the above expression. By examining the behavior of the Lanczos coefficients, one can conclude that the K-complexity and K-entropy exhibit similar behaviors to those studied in the previous subsection. The corresponding numerical results are shown in figure \ref{fig:lattice measures}. 
In the left panel, we consider the evolution of K-complexity for several values
of $z$. The same as that of continuous case, the K-complexity decreases as the dynamical exponent increases. It also exhibits
oscillations for  non-zero $\beta m$.
%one can observe oscillations in $K_\mathcal{O}(t)$. 
Beside the  oscillations, at early
times, $K_\mathcal{O}(t)$ exhibits an exponential growth corresponding to the linear behavior of the Lanczos coefficients depicted in figure \ref{fig:LClattice1}.
The slope of the curves becomes smaller for larger values of the dynamical exponent. 
Moreover, at
late times, the behavior of K-complexity is different from the exponential growth for the continuum case due to the saturation of $b_n$.

The right panel shows the evolution of K-entropy with the same values of the parameters. Again, excluding the oscillations, we see that $S_{\mathcal{O}}(t)$ exhibits a logarithmic growth that decreases as one increases $z$, which is consistent with the previous results depicted in figures \ref{fig:KVKEm0} and \ref{fig:KEntropymassive}. Furthermore, by decreasing the mass parameter the oscillations become less pronounced. Note that in these plots we have fixed $m
^z$ and therefore the period and amplitude of oscillation are approximately the same.

To close this section, note that the results shown in figure \ref{fig:lattice measures}
for $K_{\mathcal{O}}$ and $S_{\mathcal{O}}$ 
have not been computed in the previous literature including \cite{Avdoshkin:2022xuw,Camargo:2022rnt}. Indeed, in \cite{Avdoshkin:2022xuw} the authors have
just studied the scaling of $b_n$s as a function of $n$ for different values of the parameters. It is  worth mentioning, 
thanks to our  elegant numerical method discussed in sec. \ref{Lifshitz scalar theory},  we could find these measures without having an analytic expression for $\phi_0(t)$ and its higher derivatives (see eq. \eqref{phi0nth}).

\section{Conclusions}\label{Conclusions}

In this paper, we have studied the general behavior of Lanczos coefficients and K-complexity in a Lifshitz scalar field theory with nontrivial values of the dynamical critical exponent. Furthermore, we have examined the effects of mass, temperature, finite UV cutoffs in continuous momentum space, and finite lattice spacing. In the following, we summarize our main results and discuss some further problems.
\begin{itemize}
\item In a continuum massless Lifshitz scalar theory, the Lanczos coefficients grow linearly with $n$, which is consistent with the universal operator growth hypothesis. Interestingly, although the slope is completely independent of the dynamical exponent, the value of $b_n$ decreases with $z$. Hence, non-relativistic scale invariance does not influence the rate of change of the Lanczos coefficients and in particular, the staggering
behavior is also observed in this case. A curious feature that we have observed is that the staggering effect becomes less pronounced as we increase the dynamical exponent. The K-complexity exhibits exponential growth with time and decreases as $z$ is increased, although the slope at late times is the same for all values of the dynamical exponent and is given by $\frac{2\pi}{\beta}$. In this case, the K-variance stabilizes to a constant value at late times, and the fluctuations become less pronounced for larger values of the dynamical exponent. Furthermore, for K-entropy, we observe late-time linear growth with the same slope for different values of $z$.

\item In a continuum massive theory, the Lanczos coefficients exhibit qualitatively similar behavior to that of the massless case, with two interesting features. First, similar to the massless case, the slope of $b_n$ for odd and even $n$ are the same and is independent of $z$. Moreover, in the large mass regime, the separation of $b_n$ between odd and even $n$ increases as one increases the dynamical exponent and is proportional to $m^z$.

\item 
The general behavior of $K_{\mathcal{O}}(t)$ and $S_{\mathcal{O}}(t)$ is the same as that in the massless case. Specifically, while the complexity exhibits exponential growth, the entropy exhibits linear growth at late times. While the behavior is independent of $z$ at early times, the critical exponent has a significant effect at late times, such that both the complexity and entropy decrease as $z$ is increased. Furthermore, $K_{\mathcal{O}}(t)$ and $S_{\mathcal{O}}(t)$ are decreasing functions of the mass parameter. Indeed, similar to the relativistic case, we expect that the effective length of quantum fluctuations is decreased by mass, and hence both measures decrease.

\item In the presence of a hard UV cutoff, the behavior of the Lanczos coefficients is significantly modified. Initially, they exhibit a phase of linear growth, which is followed by saturation to a constant value for large $n$. As a consequence, the K-complexity exhibits exponential growth at relatively early times, followed by linear growth at late times. These phases are associated with the linear and saturation phase of the Lanczos coefficients, respectively. Furthermore, the K-entropy exhibits approximate logarithmic growth, which is related to the linear growth of complexity at late times. Similarly, when considering a discretized version of the model with a finite lattice spacing, we found similar results.

\item For a discretized version of our model with a finite lattice spacing in one spatial dimension with periodic boundary condition, we have found similar results as that  the 
case with a hard UV cutoff. Namely, the Lanczos coefficients grow linearly with $n$ and then saturate to a constant value which is decreasing as a function of the temperature. During the linear growth regime, the slope approximately remains constant; independent of the dynamical exponent.
Again, a staggering effect is observed, which decreases with increasing $z$. Similar to the continuous case the K-complexity decreases as dynamical exponent increases. Excluding the small oscillations which is due to nonzero mass, at early times, $K_{\mathcal{O}}(t)$ exhibits an exponential growth corresponding to the linear behavior of $b_n$. Moreover, at late times, growth behaviors of K-complexity are
different from the exponential growth for the continuum case due to the saturation of the Lanczos coefficients. Also $S_{\mathcal{O}}(t)$ exhibits a logarithmic growth and decreases as $z$ increases, which is consistent with the previous results.

\end{itemize}

Recall that \cite{MohammadiMozaffar:2017nri, He:2017wla} suggested that in the massless limit by increasing the dynamical exponent the theory, i.e., Eq. \eqref{Lifaction} or its discretized version, starts to show nonlocal effects such that for $z\gg 1$ it becomes highly nonlocal. Indeed, in this regime the dynamical exponent produces correlations between long distance lattice points and hence entanglement does not occur only at the boundary. In this case, we have a crossover from the area law (which happens for small $z$) to the volume law for the entanglement entropy. Our results for the K-complexity and other related quantities show that the scaling does not change for non trivial values of the dynamical exponent. Based on this observation we claim that nonlocal effects  do not change the qualitative behavior of K-complexity. As another example consider the following nonlocal scalar theory first introduced in \cite{Shiba:2013jja}
\begin{eqnarray}
\mathcal{L}= \frac12  \left[ (\frac{d\phi}{dt})^2-B_0 \phi e^{A (-\partial^2)^{\frac{w}{2}}}\phi \right].
\end{eqnarray}
The above model also exhibits volume law entanglement for the ground state as long as the size of the subsystem is smaller than a certain scale. The corresponding dispersion relation is given as follows
\begin{equation}
 \epsilon_k=\sqrt{e^{A_0(k^2)^{\frac{w}{2}}}}=e^{\frac{A}{2}(k^2)^{\frac{w}{2}}}
\end{equation}
In this model, we can find different measures in Krylov space numerically where the results show that the non-locality parameter $A$ has no effect on Lanczos coefficients and K-complexity. Indeed, the non-locality parameter appears as an overall coefficient in $f^w$ and thus its effect disappears due to the normalization condition.

Another interesting observation that can be made is that, at least over the range of our numerical computations, it appears that the behavior of the K-complexity is similar to that of the exponential of the K-entropy. Our numerical computations show that in the region where the complexity exhibits exponential growth, the K-entropy grows linearly, while when the complexity exhibits linear growth, the K-entropy has logarithmic behavior.

Actually, an alternative definition of complexity has been proposed in the context of spread complexity \cite{Balasubramanian:2022tpr} (see also \cite{Fan:2022xaa}), where the complexity is given as 'the exponential of the entropy of the probability distribution of weights in an orthonormal basis' \cite{Balasubramanian:2022tpr}: ${\cal C} = e^{\cal S}$.

Of course, in our case, we would not expect to obtain such an exact relation between complexity and entropy, as it is evident from their definitions in Eqs. \eqref{KC} and \eqref{Kentropy}, respectively. Nonetheless, from our numerical results, we have found that at least for the massless case, we have $1 + K_{\mathcal{O}}(t) \sim e^{a S_{\mathcal{O}}(t)}$ for some numerical constant $a < 1$, which is not universal. Indeed, its value depends on the dimension and critical exponent.

It would be interesting to see the precise information that could be obtained from complexity and entropy in the context of the dynamics of Krylov space. Indeed, since in this context, all information is encoded in the Lanczos coefficients, one would suspect that complexity may have additional information compared to entropy.

\section*{Acknowledgments}
We would like to thank Souvik Banerjee and Pawel Caputa for discussions on different aspects of Krylov complexity. Some numerical computations related to
this work were carried out at IPM Turin Cloud Services \cite{turin}. We would also like to thank the referee for her/his
useful comments. M.A. and M.J.V. are supported by Iran National Science Foundation (INSF) under project No. 4023620.

\end{document}